\documentclass[aps,prl,superscriptaddress,floatfix,nofootinbib,showkeys,showpacs,twocolumn,reprint]{revtex4-1}

\usepackage{aas_macros}
\usepackage{amsmath}
\usepackage{graphicx}
\usepackage[normalem]{ulem}
\usepackage{color}
\usepackage{xspace}
\newcommand{\todo}[1]{\textbf{\color{red}{#1}\color{black}}}

\renewcommand{\todo}[1]{}

\newcommand{\jcap}{JCAP}

\newcommand{\Grays}{$\gamma$~rays\xspace}
\newcommand{\gray}{$\gamma$-ray\xspace}
\newcommand{\unit}[1]{\,\mathrm{#1}\xspace}
\newcommand{\Fermi}{\emph{Fermi}\xspace}
\newcommand{\FermiLAT}{\emph{Fermi}~LAT\xspace}
\newcommand{\fermiLAT}{\emph{Fermi}-LAT\xspace}

\begin{document}


\title{The Fermi Large Area Telescope as a Galactic Supernovae Axionscope}

\author{M.~Meyer}
\email{manuel.meyer@fysik.su.se}
\affiliation{Department of Physics, Stockholm University, AlbaNova, SE-106 91 Stockholm, Sweden}
\affiliation{The Oskar Klein Centre for Cosmoparticle Physics, AlbaNova, SE-106 91 Stockholm, Sweden}
\author{M.~Giannotti}
\email{mgiannotti@barry.edu}
\affiliation{Physical Sciences, Barry University,
11300 NE 2nd Ave., Miami Shores, FL 33161, USA}
\author{A.~Mirizzi}
\email{alessandro.mirizzi@ba.infn.it}
\affiliation{Dipartimento Interateneo di Fisica ``Michelangelo Merlin'', Via Amendola 173, 70126 Bari, Italy}
\affiliation{Istituto Nazionale di Fisica Nucleare - Sezione di Bari, Via Amendola 173, 70126 Bari, Italy}
\author{J.~Conrad}
\affiliation{Department of Physics, Stockholm University, AlbaNova, SE-106 91 Stockholm, Sweden}
\affiliation{The Oskar Klein Centre for Cosmoparticle Physics, AlbaNova, SE-106 91 Stockholm, Sweden}
\affiliation{Wallenberg Academy Fellow}
\author{M.A.~S\'anchez-Conde}
\affiliation{Department of Physics, Stockholm University, AlbaNova, SE-106 91 Stockholm, Sweden}
\affiliation{The Oskar Klein Centre for Cosmoparticle Physics, AlbaNova, SE-106 91 Stockholm, Sweden}

\date{\today}

\begin{abstract}
In a Galactic core-collapse
supernova (SN), axionlike particles (ALPs) 
could be emitted
via the Primakoff process and eventually convert into \Grays
 in the magnetic field of the
Milky Way.  
From a data-driven sensitivity estimate, we find that, for a SN exploding 
in our Galaxy,
the \Fermi Large Area Telescope (LAT) would be able to explore the photon-ALP coupling down to 
$g_{a\gamma} \simeq 2 \times 10^{-13}$~GeV$^{-1}$ for an ALP mass
 $m_a \lesssim 10^{-9}$~eV. 
 These values are out of reach of next generation laboratory experiments.  
In this event, the \FermiLAT would probe large regions of the ALP parameter space invoked to explain the anomalous transparency 
of the Universe to \Grays,  
stellar cooling anomalies, and cold dark matter. 
If no $\gamma$-ray emission were to be detected,
\fermiLAT observations would improve current bounds derived from SN\,1987A by more than one order of magnitude. 
\end{abstract}



\maketitle

\emph{Introduction.}---Axionlike particles (ALPs) are light pseudo-scalar bosons with a two-photon coupling $a \gamma \gamma$ of strength $g_{a\gamma}$ which 
are predicted by several extensions of the Standard Model (see~\cite{Jaeckel:2010ni} for a review).
These particles can constitute a significant fraction or the entire cold dark matter~\cite{preskill1983,abbott1983,dine1983,marsh2011,Arias:2012az}.
In the presence of an external magnetic field, the $a \gamma \gamma$ coupling  leads to photon-ALP mixing~\cite{Raffelt:1987im}.
 This  effect is exploited to search for ALPs in light-shining-through-the-wall experiments such as ALPS~\cite{Ehret:2010mh,baehre2013}, for solar ALPs (e.g.~the CAST experiment~\cite{Arik:2011rx}) and for ALP dark matter in cavity experiments (e.g.~ADMX~\cite{Duffy:2006aa}). 
The best experimental bound on the photon-ALP coupling is $g_{\rm a\gamma}\lesssim 8.8\times 10^{-11}$~GeV$^{-1}$ obtained with the CAST experiment for 
$m_a\lesssim 0.02$~eV~\cite{Arik:2011rx}.
ALP production in stars via the Primakoff process \cite{Raffelt:1985nk} would also cause an additional energy drain that
may change the stellar lifetime, eventually beyond the limits allowed by  observations (see \cite{Carosi:2013rla} for a review).  
For masses $m_a \lesssim \mathrm{keV}$,
one finds $g_{a \gamma} \lesssim 8 \times 10^{-11}$~GeV$^{-1}$ from Cepheid stars~\cite{Friedland:2012hj}, 
and $g_{a \gamma} \lesssim 6.6  \times 10^{-11}$~GeV$^{-1}$ from globular clusters~\cite{Ayala:2014pea}.

For ALPs with masses $m_a\lesssim 10^{-9}$~eV, the strongest bound on $g_{a \gamma}$ is derived from the absence of \Grays
from SN\,1987A, a core-collapse supernova (SN) that exploded in the Large Magellanic Cloud at a distance of about 50 kpc. 
ALPs would be copiously emitted by the SN core
and part of this flux would be converted to \Grays in the Galactic magnetic field (GMF).
Upper limits on the
SN\,1987A $\gamma$-ray flux by the Gamma-Ray Spectrometer on the Solar Maximum Mission satellite imply a bound on the coupling~\cite{Brockway:1996yr,Grifols:1996id}.
 A recent analysis with state-of-the-art models, both for the GMF and for the production of ALPs in the SN, results in
$g_{a\gamma} < 5 \times 10^{-12}\unit{GeV}^{-1}$ for  $m_{a}< 10^{-10}\unit{eV}$~\cite{Payez:2014xsa}.
This bound significantly constrains the parameter space for photon-ALP conversions in 
 large-scale magnetic fields proposed as 
a mechanism to explain evidence for a reduced \gray  
absorption on the extragalactic background light~\cite{Simet:2007sa,DeAngelis:2008sk,Fairbairn:2009zi,DeAngelis:2011id,Dominguez:2011xy,Meyer:2013pny,Rubtsov2014} 
(see, however, \cite{biteau2015,dominguez2015})
as well as most of the low-mass region where ALPs could
 explain observed stellar cooling anomalies~\cite{Giannotti:2015kwo}.

The \Fermi Large Area Telescope (LAT) is sensitive to \Grays with  
 energies from 20\,MeV to $>300\,$GeV
 and monitors the entire sky every three hours.
It is therefore perfectly 
suited to search for the ALP-induced \gray burst from the next Galactic SN.
Motivated by this perspective, our work aims to 
present a detailed evaluation of the \Fermi-LAT sensitivity to the photon-ALP coupling
from a SN event.

\emph{ALP production in a supernova core.}---
ALPs would be produced in a stellar medium primarily through the Primakoff process~\cite{Raffelt:1985nk}, in which thermal photons are converted into ALPs in the electrostatic field of ions, electrons, and protons.
In order to calculate the ALP production rate in a SN core via the Primakoff process we closely follow Ref.~\cite{Payez:2014xsa}.
We find 
\begin{eqnarray}
\dfrac{d \dot n_a}{dE}&=&
\frac{g_{a\gamma}^{2}\xi^2\, T^3\,E^2}{8\pi^3\, \left( e^{E/T}-1\right) } \nonumber \\
& &\left[ \left( 1+\dfrac{\xi^2 T^2}{E^2}\right)  \ln(1+E^2/\xi^2T^2) -1 \right] \,,
\label{eq:axprod}
\end{eqnarray}
where $E$ is the photon energy, $T$ the temperature, and 
$\xi^2={\kappa^2}/{4T^2}$ with $\kappa$ the inverse Debye screening length, due 
to the finite range of electric field surrounding charged particles in the plasma.  
The total ALP production rate per unit energy is found by
 integrating Eq.~\eqref{eq:axprod} over the SN volume. 
We consider one-dimensional SN models with progenitor masses of $10$ and $18$~$M_{\odot}$ \cite{Fischer:2009af}, 
and account for the effects of partial proton degeneracy and effective nuclear mass at high density.

Integrated over the explosion time, which is of the order of 10\,s, we find that the ALP spectrum can be parametrized by a power law with exponential cutoff,
\begin{equation}
\frac{dN_a}{dE} = C \left(\frac{g_{a\gamma}}{10^{-11}\unit{GeV}^{-1}}\right)^2
\left(\frac{E}{E_0}\right)^\beta \exp\left( -\frac{(\beta + 1) E}{E_0}\right) \,\, 
\label{eq:time-int-spec}
\end{equation}
with $ C $, $ E_0 $, and $ \beta $ given in Tab.~\ref{tab:rate_parameters}.
\begin{table}[h]
	\begin{center}
		\begin{tabular}{ l  c  c   c  }
			\hline \hline
			Progenitor mass			& $ C $ [$10^{50}\unit{MeV}^{-1} $]		&  $ E_0 $ [MeV] 					& $ \beta  $    \\ \hline 
			$10\,M_\odot$	& $ 5.32$  	& $ 94 ~ $  	& 2.12		\\ 
			$18\,M_\odot$	&	$ 9.31 $	& $102 ~ $   	& 2.25  \\ \hline
		\end{tabular}
		\caption{Best fitting values for the parameters in Eq.~\eqref{eq:time-int-spec}.}
		\label{tab:rate_parameters}   
	\end{center}
\end{table}

\emph{ALP--photon conversions in the Milky Way.}---Given the 
photon-ALP couplings we are considering here, the mean free path of ALPs inside the SN is much larger than the SN radius.
Hence, once produced, they can escape the star without further interactions~\cite{Brockway:1996yr}
and subsequently convert into photons in the GMF.

\todo{
For sufficiently light ALPs and a magnetic field homogeneous over a length $l_c$, 
the photon-ALP conversion probability becomes practically energy independent for the energies considered here, 
$P_{a\gamma} \sim (g_{a\gamma} B_T l_c)^2 / 4,$
 where $B_T$ is the component of the magnetic field perpendicular to the photon beam direction.~\cite{Raffelt:1987im}. 
However, radio synchrotron observations as well as Faraday rotation measures of pulsars 
suggest that the GMF is not homogeneous but consists of both a coherent and a turbulent component~\cite{Han:2006ci,Han:2004aa}.
In the following, we neglect the turbulent component
as the typical coherence length of the GMF is much smaller 
than the photon-ALP oscillation length for the energies, magnetic field strength, 
and ALP parameters considered here (see the Supplemental Material in Ref. \cite{ajello2016}).
The regular component 
is parallel to the Galactic plane, with 
has a typical strength $B \simeq$ a few~$\mu{\rm G}$ and a radial coherence length  $l_c \simeq10~{\rm kpc}$~\cite{Beck:2008ty}. 
}

We use the coherent component of the \citeauthor{Jansson:2012pc} model~\cite{Jansson:2012pc} (JF12 hereafter) as our GMF benchmark model, and describe the Galactic electron density as proposed in Ref.~\cite{NE2001}.
The model discussed by \citeauthor{Pshirkov:2011um}~\cite{Pshirkov:2011um} (PTKN11 in the following)
 gives similar results for the GMF
in our regions of  interest. 
We closely follow the technique described in Ref.~\cite{Horns:2012kw}  to solve the full beam propagation equation along a Galactic line of sight. 

Once the ALP production rate and the Galactic ALP-photon conversion probability are known, we find the the differential \gray flux per
unit energy integrated over the explosion time arriving at Earth 
\begin{equation}
\frac{dN_\gamma}{dE}= \frac{1}{4 \pi d^2} \frac{dN_a}{dE} \times P_{a\gamma} \,,\label{eq:diffphotonflux}
\end{equation}
where $d$ is the SN distance.

\emph{Sensitivity Estimate.}--- Current neutrino detectors are expected to measure a plenitude of neutrinos from the next Galactic SN.
For instance,
 the Super-Kamiokande water Cherenkov detector should detect about $10^4$ neutrino events 
 from a SN at $d = 10\,$kpc
  (e.g., \cite{scholberg2012,Mirizzi:2015eza}).
The ALP-induced \gray signal is expected to arrive roughly simultaneously to the neutrinos and hence 
the neutrino signal would provide the required timing information to search for a coincident \gray signal (see \cite{Raffelt:1996} for a review).

The distribution of supernovae in the Galaxy must follow regions of star formation, notably in the spiral arms.  
These distributions are very broad \cite{Mirizzi:2006}
so that any distance between 2 and 20\,kpc is almost equally likely. 
For definiteness, we estimate the sensitivity of the \FermiLAT to detect
the burst from a Galactic SN at the position of the Galactic center (GC). We use actual data instead of simulations.
This approach has the advantage that we do not rely 
on the modeling of the instrumental response functions (IRFs) within the simulations.
Since the background rate and the photon-ALP conversion probability 
change with the SN position, we also consider the test cases 
of Betelgeuse and the possibility of an extragalactic SN in M\,31 (Andromeda). 
We return to these sources in the Discussion. 

We extract a random day of \fermiLAT data (July 28, 2015)
in the energy range between 50\,MeV and 500\,MeV motivated from the expected ALP spectrum.
We do not consider lower energies since the effective area of the \FermiLAT decreases rapidly at 50\,MeV.
To minimize the contamination of the Earth Limb we only use events that arrive at zenith angles $ < 80^\circ$.
As we are expecting a short burst of a duration of 10--20\,s,
we utilize events passing the 
\texttt{P8R2\_TRANSIENT020} selection cuts
which are analyzed with the corresponding \texttt{V6} IRFs.\footnote{See \url{http://www.slac.stanford.edu/exp/glast/groups/canda/lat_Performance.htm}}

For the one day of data, we calculate good time intervals (GTIs)  
for which data quality cuts are fulfilled and the region of interest is not contaminated by the \Grays from 
the Earth Limb.  
The SN will be detectable with the \FermiLAT if
it occurs at a time within one GTI, which is assumed here.
We bin each GTI in time bins of 20\,s, 
so that the entire burst would be contained in one bin. 
We extract the point spread function (PSF) at 50\,MeV 
in each bin using the \texttt{gtpsf} tool included in the \emph{Fermi Science Tools} version 10r01p01 and 
determine the 68\,\% containment radius, $r_{68}$, 
and its time average over the whole GTI, $\langle r_{68} \rangle$.\footnote{See \url{http://fermi.gsfc.nasa.gov/ssc/data/analysis/}.}
During one GTI, the PSF and exposure change slightly as the 
source moves through the field of view. 

For each GTI we generate the 20\,s light curve for $\gamma$-ray events that arrive within $\langle r_{68} \rangle$ from the GC. 
In an SN event, the time bin containing the potential
signal plus background ($x$ ``ON'' counts)
would be tagged by the detection of the neutrino events. 
The remaining time bins $i$ with counts $y_i$  (``OFF'' counts)
and relative exposure $\epsilon_i$ in comparison to the ON bin
can be used to estimate the expected number
of background counts in the ON bin, $b$.
For one GTI, we use all time bins for the background estimation 
and the exposure of the central time bin of the light curve as the ON exposure.
Maximizing the standard Poisson likelihood 
for the OFF bins, one finds the maximum likelihood estimator
$\hat{b} = \alpha y$, with $y = \sum_i y_i$ 
and $\alpha = (\sum_i\epsilon_i)^{-1}$.
Stepping through the expected number of signal counts
$\mu$ we derive 
the 95\,\% confidence interval for observed counts $x$
using the method of \citeauthor{fc1998} \cite{fc1998}.
Assuming no SN signal, we set $x$ to be equal to the smallest 
integer greater than or equal to $\hat{b}$, 
i.e. $x = \lceil \hat{b} \rceil$, and calculate  
the 95\,\% upper limit on the number of expected counts, $\mu_\mathrm{UL}$, 
from the confidence interval. 
The method of setting the observed counts equal to the expected number of counts is often referred to as an ``Asimov data set'' \cite{cowan2011}.

In the following, we focus on one representative GTI that gives the median value for $\mu_\mathrm{UL}$ for all considered GTIs.
We list the values for the start time $t_0$,
the length of the considered GTI $\Delta t$,
 as well as  $\alpha$, $\hat{b}$, and $\mu_\mathrm{UL}$  in Tab. \ref{tab:src}.
 
\begin{table}[tbh]
\centering
\begin{tabular}{lcccc}
\hline
\hline
{} & GC & Betelgeuse & M\,31 & SN\,1987A\\
\hline
R.A. $(^\circ)$ & $266.42$ & $88.79$ & $10.63$ & $83.87$ \\ 
Dec. $(^\circ)$ & $-28.99$ & $7.41$ & $41.30$ & $-69.27$\\
Distance (kpc) & $8.5$ & $0.197$ & $778$ & 51.4\\
\hline
$t_0$ (MJD) &  $57,231.582$& $57,231.284$ & $57,231.144$ & $54,757.806$\\
$\Delta t$ (s) & $1581$& $1519$ & $1079$ & $867$ \\
\hline
$\langle r_{68} \rangle (^\circ)$ & 10.92& 9.73 &10.37 & 8.94\\
$\hat{b}$ & 3.32 & 1.11 & 0.94 & 1.05\\
$\alpha$ & 0.014 & 0.014 & 0.030 & 0.024\\
$\mu_\mathrm{UL}$ & 6.43 & 5.61 & 4.19 & 5.67\\ 
\hline
\end{tabular}
\caption{\label{tab:src}Positions and times 
for the data extraction  
for the considered hypothetical SN.
Also listed are the 68\,\% containment radius time-averaged over the  GTI ($\langle r_{68} \rangle$), the expected background counts ($\hat{b}$), relative exposures ($\alpha$), and upper limits on  signal counts ($\mu_\mathrm{UL}$).}

\end{table}

To translate this upper limit into an expected limit on the ALP parameters,
we multiply the time-integrated ALP spectrum (cf. Eq.~\eqref{eq:time-int-spec}) 
with the photon-ALP conversion probability, integrate it over the considered energy range,
 and fold it with the \fermiLAT IRF.
The IRF is generated with the \texttt{gtrspgen} tool for every time bin of the light curve.
To account for the PSF, we have to multiply this number with the fraction
of counts contained within $\langle r_{68} \rangle$ ($\approx 0.68$ at 50\,MeV). 
For energies above 50\,MeV this fraction becomes 
$> 0.68$ as the PSF improves. 
However, the sensitivity is dominated by the lowest energies 
 due to the shape of the ALP spectrum.
We therefore make the assumption that this fraction is equal to 0.68 
over the entire energy range. 
We have checked that including the full PSF has a negligible effect on the final sensitivity.
We consider ALP parameters on a logarithmically spaced $(24 \times 24)$ grid in
ALP mass and photon-ALP coupling with $-3 \leq \log_{10}(g_{11}) \leq 1$ and 
$-1 \leq \log_{10}(m_\mathrm{neV}) \leq 3$,
where $g_{11} = g_{a\gamma} / 10^{-11}\unit{GeV}^{-1}$
and $m_\mathrm{neV} = m_a / \mathrm{neV}$.
For a fixed ALP mass, we find $\mu \propto g_{a\gamma}^4$, as expected since both the ALP production and the 
photon-ALP conversion scale as $g_{a\gamma}^2$.   
ALP parameters that result in $\mu \geq \mu_\mathrm{UL}$ 
can be excluded at $95\,\%$ confidence.

\begin{figure*}[thb]
\centering
\includegraphics[width = .95\linewidth]{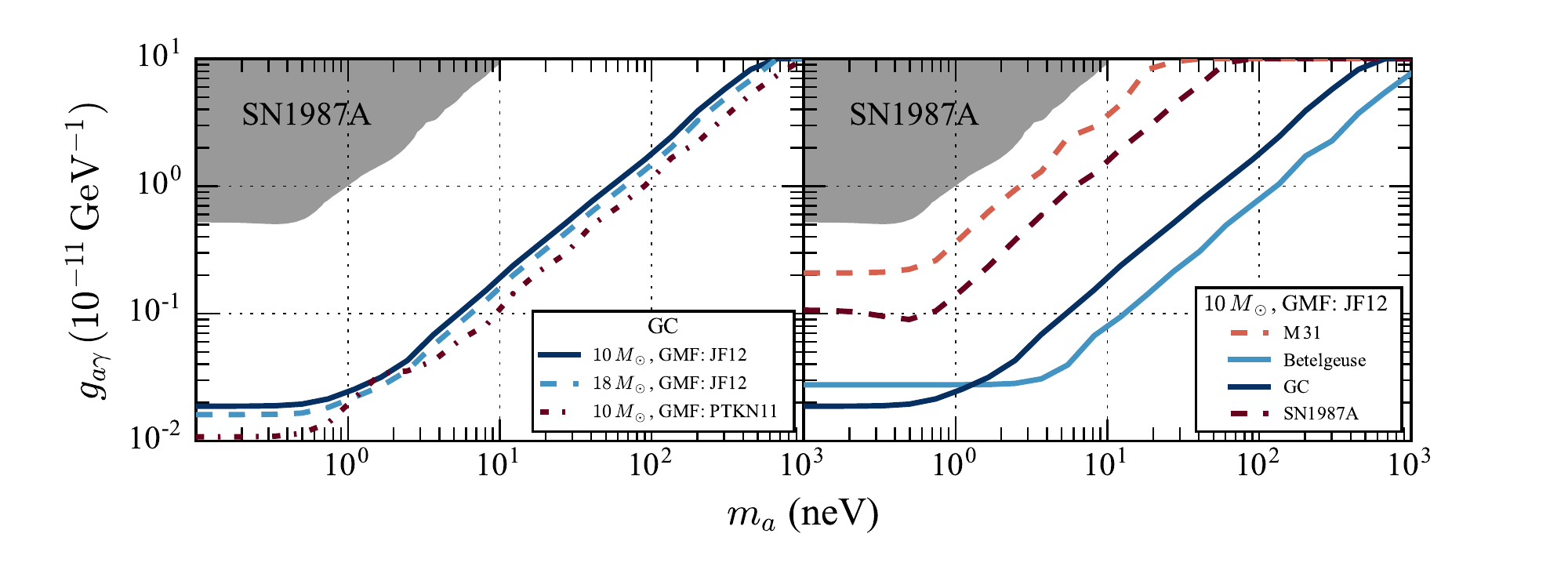}
\caption{\label{fig:sys}
Expected
 limits on ALP parameters from a SN explosion.
The gray shaded region shows the constraints from SN\,1987A.
\textit{Left}: Limits for different progenitor masses and GMF models.
\textit{Right}: Limits for different potential SN positions.}
\end{figure*}

\emph{Discussion.}---We show the expected upper limit on the ALP parameters 
in the left panel of Fig. \ref{fig:sys} for two GMF models (JF12 and PTKN11) and the two progenitor masses $10\,M_\odot$ and $18\,M_\odot$. 
Depending on the time bin in which the SN explosion occurs,
the exposure of the \FermiLAT varies, resulting in 
an uncertainty of a factor up to $\sim 1.6$ for each GTI. 
The solid lines depict the median limit values. 
The upturn of the limits around $m_a = 1\,\mathrm{neV}$ 
is due to a reduced conversion probability for high ALP masses. 
In the absence of a signal, regardless of the progenitor mass and the GMF, 
the \fermiLAT observations of a Galactic SN would improve the current SN\,1987A limit~\cite{Payez:2014xsa} (gray shaded region in Fig.~\ref{fig:sys}) by over an order of magnitude.

The number of expected counts increases by $\sim75\,\%$ for the highest progenitor mass considered, yet the $g_{a\gamma}^4$ dependence of $\mu$ leads only 
to a marginal improvement of the limits by a factor of $(1.75)^{1/4} \sim 1.15$.
Hence our limits are nearly insensitive to the exact progenitor mass. 
We will restrict 
ourselves to $10\,M_\odot$ progenitors in the following.
Stars with masses $< 10\,M_\odot$ might also 
explode as an SN, however the exact minimum value 
remains a topic of on-going discussion 
and would only minimally change our results \cite[e.g.][]{jones2013,woosley2015}. 

Using the PTKN11 GMF model instead of JF12 improves the limits by almost a factor of two (73\,\%) at $m_\mathrm{neV} = 0.1$ (red solid line  in the left panel of Fig. \ref{fig:sys}), making
 the JF12 model a conservative choice. 
 Yet, one should note that the conversion probability depends on the sky position and distance of the SN 
and it is not expected that the PTKN11 would always result in better limits.

We perform the same analysis for a total of ten GTIs on the same day
 and for a GTI at another date 
($t_0 = 55,153.989$\,MJD) and find the change of the limits to be negligible. 
We furthermore test how the binning affects the limits by changing the bin size of the light curve to 
30\,s and 60\,s, respectively. 
The longer integration time leads to higher background rates 
which result in upper limit values $\mu_\mathrm{UL}  = 7.25$ and $8.37$, respectively, against the $6.43$ found in our fiducial analysis. 
We find decreased limits 
on the photon-ALP coupling by $3\,\%$ and $6\,\%$, respectively. 
We conclude that the major systematic uncertainty of our analysis is related to the choice of the GMF model. 

As a further example, we investigate the expected limits for a SN of  the red supergiant Betelgeuse (see Tab. \ref{tab:src}).
We assume the JF12 GMF model and a 10\,$M_\odot$ progenitor mass which is within the mass estimate of 
 \citeauthor{neilson2011} who find $11.6^{+5.0}_{-3.9}\,M_\odot$  \cite{neilson2011}.
The small distance of $\sim197\,$pc implies a low conversion probability in the GMF.
This is compensated by the higher flux relative to a SN in the GC.
Furthermore, the expected number of background counts $\hat{b}$ is 
 about three times smaller for the region of interest around
Betelgeuse compared to the Galactic center (cf. Tab. \ref{tab:src}). 
These points eventually lead to similar limits 
(cf. Fig. \ref{fig:sys}).
Using a local measurement of the $B$ field close to the line of sight to Betelgeuse \cite{harvey-smith2011} leads to a substantial improvement of the limits by a factor of 4.5,
making the JF12 model again a conservative choice.
We also calculate the expected limit for an extragalactic SN 
in M\,31 
at a distance of $\sim 778\,$kpc. 
The limit would improve compared to SN\,1987A constraints by a factor of $\sim2$,
also thanks to the low number of background counts expected in this direction.
 (cf. Tab. \ref{tab:src}
and Fig. \ref{fig:sys}). 
Including the possibility of ALP-photon conversions in the magnetic field of M\,31 (with a strength of 5\,$\mu$G, coherent over 20\,kpc \cite{conlon2014}) improves the limits substantially by a factor of 5. 
However, the main challenge would be the detection of a neutrino signal from a SN in M\,31. 
Super-Kamiokande could detect one neutrino event which could be connected to a SN
if an optical counterpart was found within a day of the explosion \cite{ando2005}. 
In general, it will take the next generation of Mton class water Cerenkov neutrino detectors to reliably detect 
extragalactic SN with possibly 0.1 neutrino events per year and up to $100$ events
from a SN in M\,31 \cite{ando2005,kistler2011}. 
Until then, a source stacking over longer integration times  for many extragalactic supernovae, 
similar to the analysis of Ref. \cite{ackerman2015:sn}, might be a possible venue. This is left for future work. 

We also repeat the analysis for a 
position coincident with SN\,1987A and find that current limits would improve by a factor of $\sim5$ (dark-red dashed line in the right panel of Fig. \ref{fig:sys}).  

\begin{figure}[thb]
\centering
\includegraphics[width = .95\linewidth]{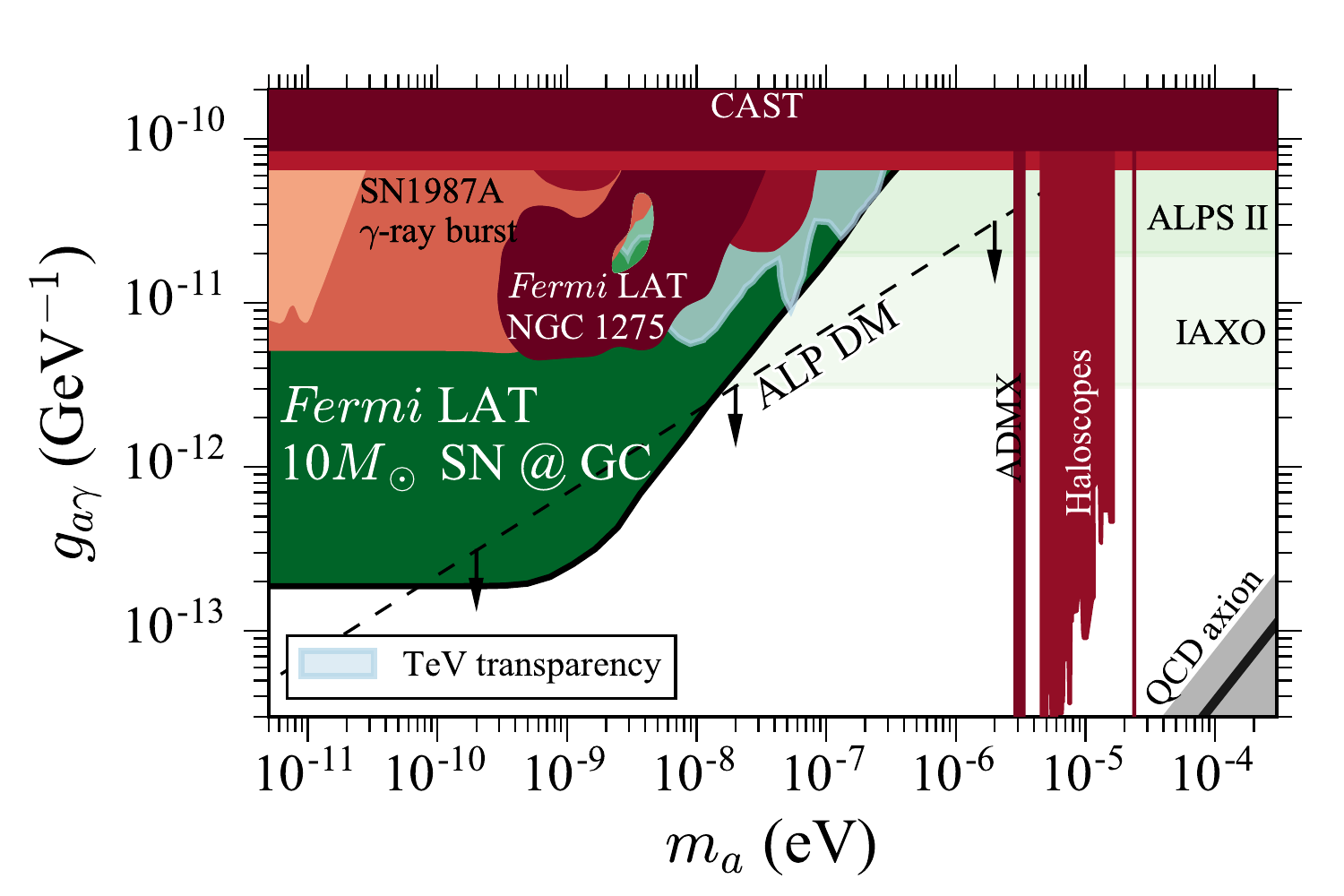}
\caption{\label{fig:limits}Comparison of the expected \fermiLAT sensitivity  (dark green) with limits (red)
and sensitivities of future experiments (green). The QCD axion band is shown in gray. 
The transparency anomaly hint could be explained by ALPs
within the blue shaded region.
ALPs with parameters below the dashed line could constitute the entire dark matter.}
\end{figure}

\emph{Conclusions.}---We compare the expected ALP limits from a SN of a $10\,M_\odot$ progenitor in the GC calculated 
with the JF12 GMF model with other limits and sensitivity projections 
in Fig. \ref{fig:limits}.
A SN event within the lifetime of the \FermiLAT 
would allow an unprecedented exploration of the ALP parameter space for ALP masses 
below 100\,neV, surpassing current bounds \cite{hess2013:alps,wouters2013,Payez:2014xsa,ajello2016,berg2016arXiv}
and the projected sensitivity of future dedicated laboratory searches such as ALPS II \cite{baehre2013} and IAXO \cite{irastorza2011} 
for masses up to 10\,neV.
It would also be possible
 to probe portions of the so-far unconstrained
parameter space where ALPs with masses $0.01\lesssim m_\mathrm{neV}\lesssim 10$
could constitute the entire dark matter (black dashed line in Fig. \ref{fig:limits} \cite{arias2012}). 
Furthermore, such an event would provide
 a definitive verdict on the role of ALPs in explaining 
 hints of an anomalous transparency of the Universe to very-high-energy photons \cite{Simet:2007sa,DeAngelis:2008sk,Fairbairn:2009zi,DeAngelis:2011id,Dominguez:2011xy,Meyer:2013pny,Rubtsov2014} and could indicate if low mass ALPs are responsible for the additional cooling observed in different stellar systems~\cite{Giannotti:2015kwo}.

In light of this potential, the question arises how probable it is that the \FermiLAT will 
observe a Galactic SN within its remaining lifetime.
The Galactic SN rate has been estimated 
to be roughly $2{-}3$ per century \cite[e.g.][]{diehl2006,keane2008,adams2013}. 
Under the assumption that SN explosions occur as a Poisson process,
the chance for one or more supernovae to occur in one year is $\sim2{-}3\,\%$.
The \FermiLAT observes $\sim20\,\%$ of the full sky at any given 
moment. 
Assuming for instance a total lifetime of the \Fermi mission of 15\,years
this results in a $\sim2.4{-}3.5\,\%$ chance to observe at least one such event in the next 7\,years of the mission.
In case of a close SN, $d < 2\,$kpc, neutrino detectors 
could measure a signal from the Silicon burning pre-SN phase \cite{Odrzywolek:2004}
and \fermiLAT target of opportunity observations could 
further increase the detection possibility. 
The Gamma-ray Burst Monitor (GBM) on-board the \Fermi satellite
observes \Grays up to $40\,$MeV over the whole sky 
not occulted by Earth and has therefore a higher chance to observe
the next Galactic SN. A sensitivity study for the GBM is left for future work. 
Despite the 
low observation probability, 
our analysis demonstrates that the next Galactic SN 
will not only shed light on the SN explosion mechanism but could 
also be used as a powerful probe of fundamental physics. 
Future \gray missions such as e-ASTROGAM \cite{Tatischeff:2016ykb},
ComPair \cite{moiseev2015}, or PANGU \cite{wu2014}
are planned to have a high sensitivity to $\gamma$ rays at tens of MeV and are expected to have an improved angular resolution
of $\sim1^\circ$ at 100\,MeV (68\,\% containment radius).
Given their foreseen large fields of view, similar to the one of the  
\FermiLAT, such missions will be 
well suited to search
for an ALP-induced \gray burst from a Galactic or even extragalactic SN.

\begin{acknowledgments}
\section{Acknowledgments}
The work of A.M. is supported by the Italian Ministero dell'Istruzione, Universit\`a e Ricerca (MIUR) and Istituto Nazionale
di Fisica Nucleare (INFN) through the ``Theoretical Astroparticle Physics'' projects.
MASC is a Wenner-Gren Fellow and acknowledges the support of the Wenner-Gren Foundations to develop his research. 
The authors would like to thank Luca Baldini, Andrea Albert, 
Brandon Anderson, Jeremy S. Perkins, and David J. Thompson for helpful 
comments on the manuscript.
The \textit{Fermi}-LAT Collaboration acknowledges support for LAT development, operation and data analysis from NASA and DOE (United States), CEA/Irfu and IN2P3/CNRS (France), ASI and INFN (Italy), MEXT, KEK, and JAXA (Japan), and the K.A.~Wallenberg Foundation, the Swedish Research Council and the National Space Board (Sweden). Science analysis support in the operations phase from INAF (Italy) and CNES (France) is also gratefully acknowledged.
\end{acknowledgments}

\bibliographystyle{apsrev4-1}
\bibliography{fermi_sn_alp}

\end{document}